\documentclass[aps,twocolumn,prl,amsmath,amssymb,superscriptaddress,footinbib]{revtex4-1}
\usepackage{graphicx,color,multirow,bm,braket}
\usepackage[colorlinks,allcolors=black]{hyperref}

\def\STO{SrTiO$_3$}

\begin{document}

\title{Quantum paraelectricity and structural phase transitions 
in strontium titanate beyond density-functional theory}

\author{Carla Verdi}
\email{carla.verdi@univie.ac.at}
\affiliation{University of Vienna, Faculty of Physics, Computational Materials Physics, Kolingasse 14-16, 1090 Vienna, Austria}
\altaffiliation{Present address: School of Physics, The University of Sydney, 
New South Wales, 2006, Australia}
\author{Luigi Ranalli}
\affiliation{University of Vienna, Faculty of Physics, Computational Materials Physics, Kolingasse 14-16, 1090 Vienna, Austria}
\author{Cesare Franchini}
\affiliation{University of Vienna, Faculty of Physics, Computational Materials Physics, Kolingasse 14-16, 1090 Vienna, Austria}
\affiliation{Department of Physics and Astronomy, Alma Mater 
Studiorum - Universit\`a di Bologna, Bologna, Italy} 
\author{Georg Kresse}
\affiliation{University of Vienna, Faculty of Physics, Computational Materials Physics, Kolingasse 14-16, 1090 Vienna, Austria}
\affiliation{VASP Software GmbH, Sensengasse 8, 1090 Vienna, Austria}

\begin{abstract}
We demonstrate an approach for calculating temperature-dependent 
quantum and anharmonic effects with beyond density-functional theory 
accuracy. By combining machine-learned potentials and the stochastic 
self-consistent harmonic approximation, we investigate the cubic to 
tetragonal transition in strontium titanate and show 
that the paraelectric phase is stabilized by anharmonic quantum 
fluctuations. We find that a quantitative understanding of the quantum 
paraelectric behavior requires a higher-level treatment of electronic 
correlation effects via the random phase approximation. This approach 
enables detailed studies of emergent properties in strongly anharmonic 
materials beyond density-functional theory.
\end{abstract}

\maketitle

Perovskite oxides are one of the most versatile classes of materials, 
displaying a huge array of properties such as 
piezoelectricity, ferromagnetism, ferroelectricity and multiferroicity, 
as well as metal-insulator transitions and superconductivity. These 
fascinating properties 
are linked to several technological applications, ranging from 
superconductors to catalysis, thermoelectric processes, and 
nanoelectronics~\cite{Cohen1964,Babushkina1998,Guo2000,
Ohta2007,Bibes2009,Hwang2017,Spaldin2019}. 
A key role is played by structural instabilities that may distort the 
ideal cubic perovskite structure, resulting in a rich structural phase 
diagram and property landscape. Crucially, structural instabilities are 
also found to compete with ferroelectricity and may thus suppress the 
latter~\cite{Vanderbilt1995,Benedek2013}. These phenomena are governed 
by lattice quantum anharmonicity, that is the presence of higher-order 
interactions on top of the harmonic vibrations of the crystal. 

Strontium titanate (\STO) is a paradigmatic example of a perovskite 
oxide with strongly anharmonic lattice dynamics, 
and is a ubiquitous playground for 
emergent phenomena in complex oxides and heterostructures 
such as two-dimensional (2D) electron gases, polaronic properties, and 
dilute superconductivity~\cite{Cohen1964,Mannhart2007,
SantanderSyro2011,Chen2015,Behnia2019}. It famously undergoes an 
antiferrodistortive (AFD) transition from a cubic to a tetragonal 
structure below 105~K~\cite{Shirane1969}, with high proximity to 
ferroelectricity. In \STO, ferroelectricity is only incipient, meaning 
that the structure is found to remain paraelectric down to zero 
kelvin, which is attributed to quantum fluctuations~\cite{Muller1979}. Yet, 
this effect can easily be overcome by small perturbations such as strain 
or isotope substitution~\cite{Schlom2004,Itoh1999}. 
The importance of quantum paraelectricity and its tunability has been 
recognized in relation to the exotic superconducting and 
transport behavior of \STO\ at low carrier doping~\cite{Saxena2014, 
Behnia2017,Maslov2021}, with the ferroelectric (FE) soft mode fluctuations 
likely playing a key role 
in providing the pairing mechanism for superconductivity in the 
quantum critical regime~\cite{Gastiasoro2020,Feigelman2021}. 
A detailed quantitative understanding of the ground-state properties of 
quantum paraelectric materials such as \STO, as well as their temperature 
dependence, thus underpins both the study of exotic physics and the 
advancement of technological applications~\cite{Chandra2017,Levy2018}. 

First-principles methods based on density-functional theory (DFT) 
have been extensively employed to investigate the strength of the AFD 
and FE instabilities in \STO\ and the influence of strain or doping 
at 0~K and within the harmonic approximation~\cite{Vanderbilt2000,
Spaldin2014,Balatsky2015}. 
An accurate microscopic description of temperature-dependent 
anharmonic processes beyond simple approximations, 
however, poses severe challenges. 
While some approaches exist~\cite{Souvatzis2009,Hellman2011,Errea2013,
Tadano2015,Mingo2021,Zacharias2022}, they are either prohibitively 
expensive beyond simple materials or lack a general and consistent 
procedure to capture all quantum and anharmonic effects to high orders. 
For these reasons, atomistic studies of quantum fluctuations 
are generally limited to path-integral Monte Carlo 
calculations using \textit{ad hoc} parametrized effective 
Hamiltonians~\cite{Vanderbilt1996,Vanderbilt2002} 
or to solving a one-dimensional (1D) or 2D lattice-nuclear Schr\"odinger 
equation from first principles, 
thereby neglecting inter-mode phonon couplings and structural changes 
for all internal and lattice degrees of freedom~\cite{Rubio2021,
Spaldin2022}. 
Finally, it is still unknown how accurately DFT can describe these 
subtle anharmonic processes, 
which are strongly functional dependent~\cite{Kresse2008,
Spaldin2014,Rubio2021}. 

In this Letter, we propose a general framework to investigate the 
anharmonic properties of this class of materials and their temperature 
dependence with beyond-DFT accuracy, and we demonstrate it for \STO. 
We employ the stochastic self-consistent harmonic approximation 
(SSCHA)~\cite{Monacelli2021} 
in combination with machine-learned interatomic potentials, or 
force fields (MLFFs)~\cite{Aspuru-Guzik2021}. We show that this 
method seamlessly enables us to fully capture strong anharmonicities 
while retaining first-principles accuracy, and further unlocks the 
possibility to perform many-body calculations beyond DFT. Here we train 
an MLFF based on the random-phase approximation (RPA)~\cite{Ren2012} 
following the principles of $\Delta$-machine 
learning~\cite{Lilienfeld2015}. 
We analyze the effects of anharmonicity in renormalizing the phonon 
frequencies, lattice parameters and internal degrees of freedom, and 
we characterize the displacive transition between the cubic and 
tetragonal phase driven by the collapse of the zone-boundary AFD 
instability. All employed exchange-correlation functionals find that 
the paraelectric phase is stabilized at 0~K by anharmonic quantum 
fluctuations, but only the RPA delivers accurate quantitative 
agreement with experimental data. 

\begin{figure*} \begin{center}
 \includegraphics[width=\textwidth]{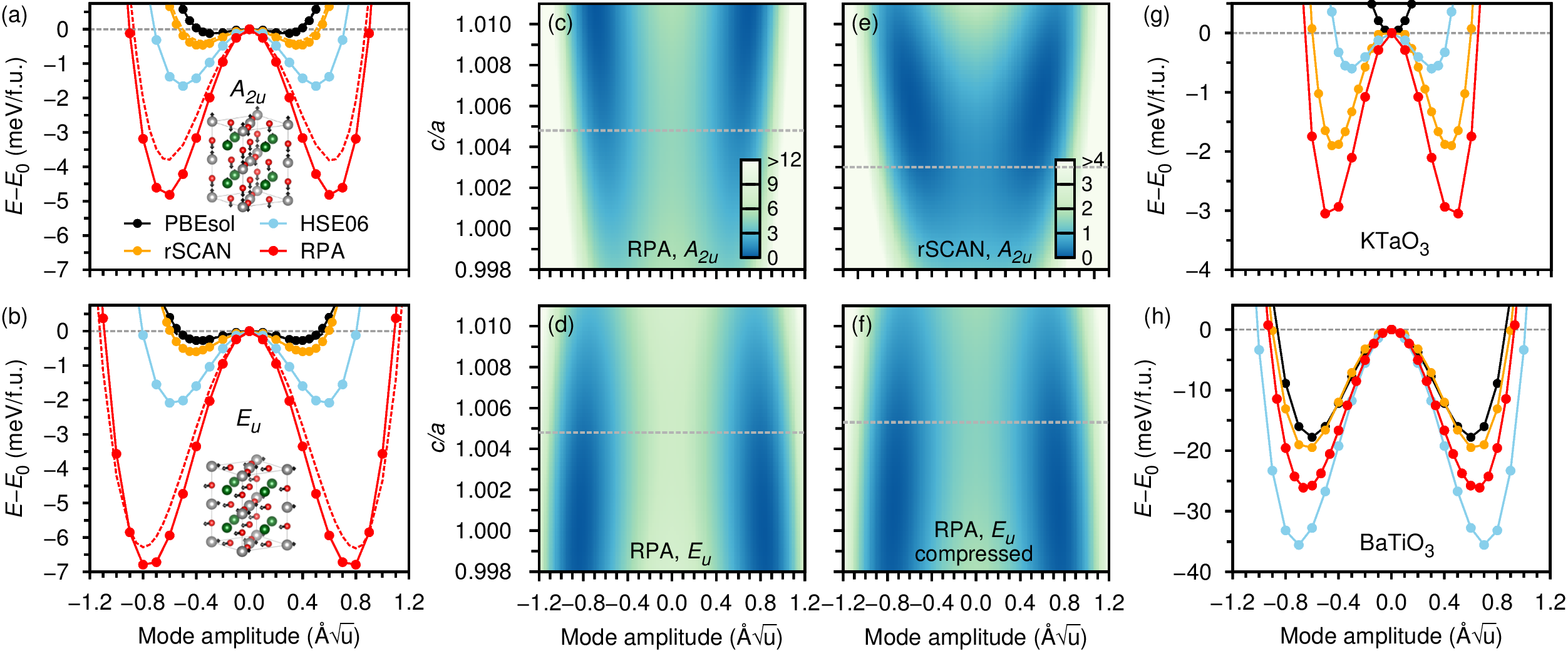}
 \caption{ \label{fig1}
  Ferroelectric potential energy surfaces (PES). 
  (a), (b) 1D PES of the $A_{2u}$ and $E_u$ FE soft modes in tetragonal 
  \STO, respectively, depicted in the insets. Results obtained using the 
  PBEsol, rSCAN and HSE06 functionals are shown, as well as the RPA. 
  The dashed lines are for the MLFFs trained on PBEsol, rSCAN (not visible) 
  and the RPA. 
  (c), (d) Corresponding 2D PES calculated at the equilibrium volume using 
  the RPA-based MLFF as a function of FE mode amplitude and $c/a$ ratio.
  (e) 2D PES for atomic displacements along the $A_{2u}$ mode 
  using rSCAN. (f) 2D PES for the $E_u$ mode calculated 
  from the RPA, after compressing the volume by 0.75\%. 
  The color maps indicate energies in meV/f.u, and the color map in 
  (c) is for all the RPA-based calculations [(c), (d), and (f)]. The 
  horizontal lines mark the equilibrium $c/a$ values. 
  (g),(h) 1D ferroelectric PES calculated for (g) KTaO$_3$ and (h) 
  BaTiO$_3$ in the cubic phase. In (g), the orange points are obtained 
  using SCAN.}
\end{center} \end{figure*}

We start by outlining our strategy for computing anharmonic properties. 
We use the SSCHA method~\cite{Errea2013,Monacelli2021}, a stochastic 
approach to the self-consistent harmonic 
approximation~\cite{Hooton1955,Koehler1966}, which 
allows us to treat anharmonicity at a nonperturbative 
level, including nuclear quantum effects. 
Through a stochastic minimization of the free energy functional, the 
SSCHA yields a variational approximation for the quantum anharmonic 
ground state of the system. 
This method delivers highly accurate data~\cite{Errea2013,Errea2019, 
Monacelli2021b}, but its applicability to large systems is limited by 
the need for computing \textit{ab~initio} energies and forces for 
several thousand supercells 
[see the Supplemental Material (SM)~\setcounter{footnote}{0}
\footnote{See Supplemental Material at [URL], which includes 
Refs.~\cite{Kaltak2014,Harl2010,Jinnouchi2020,Bartok2010,
Troester2022,Bianco2017,Cao2000,Okazaki1973}, 
for details of the first-principles calculations, the MLFFs training 
and validation, the SSCHA method and calculations, and Figs.~S1--S12.} 
for additional information]. 
A natural strategy to reduce this computational cost 
is turning to surrogate models to compute the potential energy surface 
and its derivatives. The ideal solution is offered by machine-learned 
potentials. MLFFs provide fully flexible models allowing to 
simultaneously predict the energies, atomic forces and stress 
tensor components of a given system orders of magnitude faster than 
a standard \textit{ab~initio} calculation, albeit retaining almost 
the same level of accuracy~\cite{Deringer2019,Aspuru-Guzik2021}. 
Notably, they can accurately capture the harmonic lattice dynamics as 
well as the anharmonic higher-order contributions, including the coupling 
of phonons and lattice distortions, without the need of \textit{ad~hoc} 
parametrizations~\cite{Shapeev2020,Verdi2021}. They are capable of 
describing different structural phases at the same time over a wide range 
of temperatures, and are not overly sensitive to the training data set. 
In contrast, empirical interatomic potentials, effective Hamiltonians, 
or so-called second-principles parametrizations are not able to 
simultaneously satisfy all these demands~\cite{Vanderbilt2003,
Iniguez2017,Rappe2019}.  
The combined SSCHA-MLFF approach outlined is orders of magnitude faster 
than path-integral based methods, especially in the low-temperature 
regime, and has the advantage that anharmonic phonon frequencies 
are easily obtained.

To start, we train a kernel-based MLFF for \STO\ from a database 
of DFT calculations adopting the PBEsol functional~\cite{Perdew2008}, 
using the VASP package~\cite{Kresse1993,Kresse1996,Jinnouchi2019b}. 
The MLFF is trained on the fly during molecular dynamics simulations, 
where the selection of new structures is controlled by Bayesian 
error estimation~\cite{Jinnouchi2019b}. More details 
are given in the SM~\cite{Note1}. 
The training dataset consists of 626 structures of 320 atoms each, 
sampling the configurational phase space of \STO\ up to 350~K. The 
root-mean-square errors in the energies, forces and stress 
tensors predicted by the MLFF for a test dataset are 0.18~meV/atom, 
0.037~eV/\AA\ and 0.32~kbar, respectively~\cite{Note1}. We highlight 
that attaining such small errors is crucial in the present paper, where 
energy differences of less than 1~meV/atom due to the interplay of 
different instabilities ought to be captured, as will be shown in the following. 

We first investigate the FE instability by looking at the 1D 
potential energy surface obtained by displacing the atoms in the 
unit cell along the FE soft modes driving the instability, as 
shown in Figs.~\ref{fig1}(a) and \ref{fig1}(b). In the tetragonal structure 
(illustrated in SM Fig.~S1~\cite{Note1}), the FE instability is 
split into a mode polarized parallel to the tetragonal axis $c$, 
$A_{2u}$ [Fig.~\ref{fig1}(a)] and a doubly degenerate one perpendicular to 
it, $E_u$ [Fig.~\ref{fig1}(b)]. Both modes are imaginary in standard harmonic 
calculations, as reflected by the double-well shape of the energy 
curves. The well depth of $E_u$ is larger than $A_{2u}$, however, 
in both cases it is very shallow (less than 0.5~meV per formula 
unit, f.u.), suggesting that quantum fluctuations can easily 
overcome the energy barrier. Note that our MLFF reproduces precisely 
this instability. We then perform calculations using the rSCAN 
meta-GGA functional~\cite{Yates2019} as well as the HSE06 hybrid 
functional~\cite{HSE2003,HSE2006}. As seen in Figs.~\ref{fig1}(a) 
and \ref{fig1}(b), rSCAN slightly increases the 
barriers to around 0.5~meV/f.u., and HSE06 yields even stronger 
FE instabilities. Similar calculations performed on two other 
systems, the quantum paraelectric KTaO$_3$ and the ferroelectric 
BaTiO$_3$, also show a marked dependence of the FE instability on 
the chosen density functional [Fig.~\ref{fig1}(g) and \ref{fig1}(h)].

To cure this strong functional dependence, we seek to go beyond 
semilocal and hybrid DFT by adopting the accurate, but computationally  
costly, many-body RPA method~\cite{Ren2012,Schimka2010}. 
RPA total energy calculations are at least two orders of magnitude 
more expensive than standard DFT ones, or three when forces are also 
computed~\cite{Ramberger2017}. To accelerate RPA calculations via 
machine learning, we train an RPA-based MLFF using the principles of 
$\Delta$-learning~\cite{Lilienfeld2015,Liu2022}. As described in the 
SM~\cite{Note1} in more detail, we compute the RPA corrections 
to the DFT energies and forces for a reduced set of structures. 
By training an MLFF that accurately reproduces these corrections, 
we can predict RPA-level energies and forces for any given structure. 
Incidentally, we use a similar procedure to also obtain an MLFF that 
reproduces the rSCAN potential energy surface. This yields low 
RMSEs comparable to the ones obtained for PBEsol, and excellent 
predictions for the phonon frequencies and the potential energy 
surfaces of the FE modes~\cite{Note1}. The latter are shown in 
Figs.~\ref{fig1}(a) and \ref{fig1}(b). We see that the RPA has 
a dramatic effect on the FE instability, making the energy 
barrier as high as almost 7~meV/f.u. Also in the case of KTaO$_3$ 
[Fig.~\ref{fig1}(g)] and BaTiO$_3$ [Fig.~\ref{fig1}(h)] 
the RPA increases the energy barrier, though for BaTiO$_3$ the well 
depth is largest using HSE06.

To gain more insight into the strength of the FE instability as 
described by different approximations to the exchange and correlation, 
we consider its behavior as a function of volume size and shape in 
\STO. To illustrate this, we compute a 2D energy 
map by displacing the atoms from their optimized tetragonal positions 
along the FE modes and varying the values of the tetragonal distortion 
$c/a$. Figures~\ref{fig1}(c) and \ref{fig1}(d) show the results 
obtained for the $A_{2u}$ 
and $E_u$ modes, respectively, at the equilibrium volume using our 
RPA-based MLFF, while in Fig.~\ref{fig1}(f) a volume 0.75\% smaller 
than the equilibrium one is used for the $E_u$ mode. 
After inspecting these plots, we can conclude that (i) both the $E_u$ 
and $A_{2u}$ instabilities display a strong dependence on strain, but 
they behave in opposite ways as a function of the lattice elongation. 
The $A_{2u}$ instability becomes stronger with increasing $c/a$ 
values and the $E_u$ one gets weaker, while they both decrease as 
a function of volume (consistent results are obtained for volume 
expansion, see SM Figs.~S6 and S7~\cite{Note1}). Moreover, (ii) the 
RPA description of the electronic structure generally yields much 
stronger FE instabilities than rSCAN [Fig.~\ref{fig1}(e)] and PBEsol (SM 
Fig.~S5~\cite{Note1}), even when considering the effects of strain. 
These observations have important implications when considering 
anharmonic fluctuations and the coupling with lattice distortions, as 
we will see in the following. 

\begin{figure}[b] \begin{center}
 \includegraphics[width=\columnwidth]{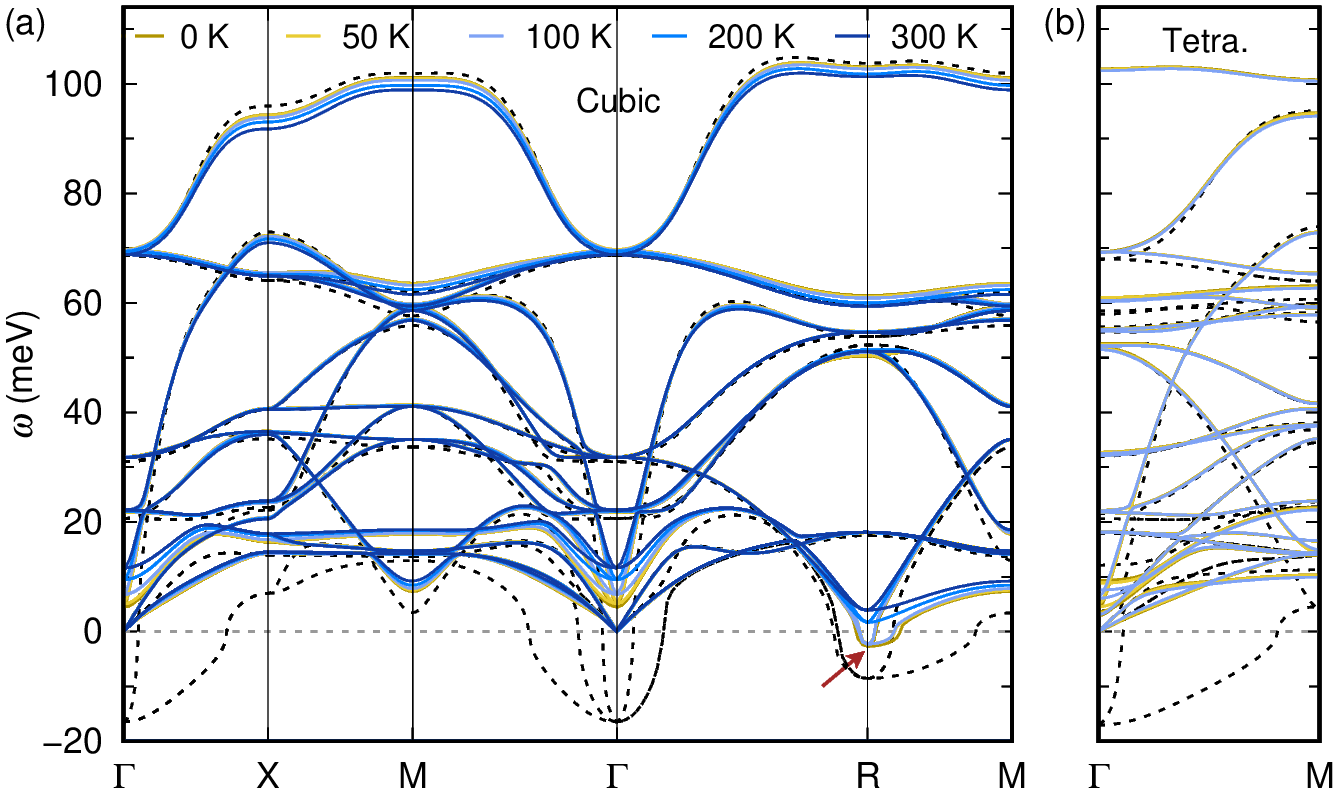}
 \caption{ \label{fig2}
 Temperature-dependent phonon dispersions of (a) cubic and 
 (b) tetragonal \STO\ calculated using the SSCHA and the MLFF trained 
 on the RPA. The dashed lines are the harmonic results, and 
 negative values denote imaginary frequencies. } 
\end{center} \end{figure}

\begin{figure*} \begin{center}
 \includegraphics[width=\textwidth]{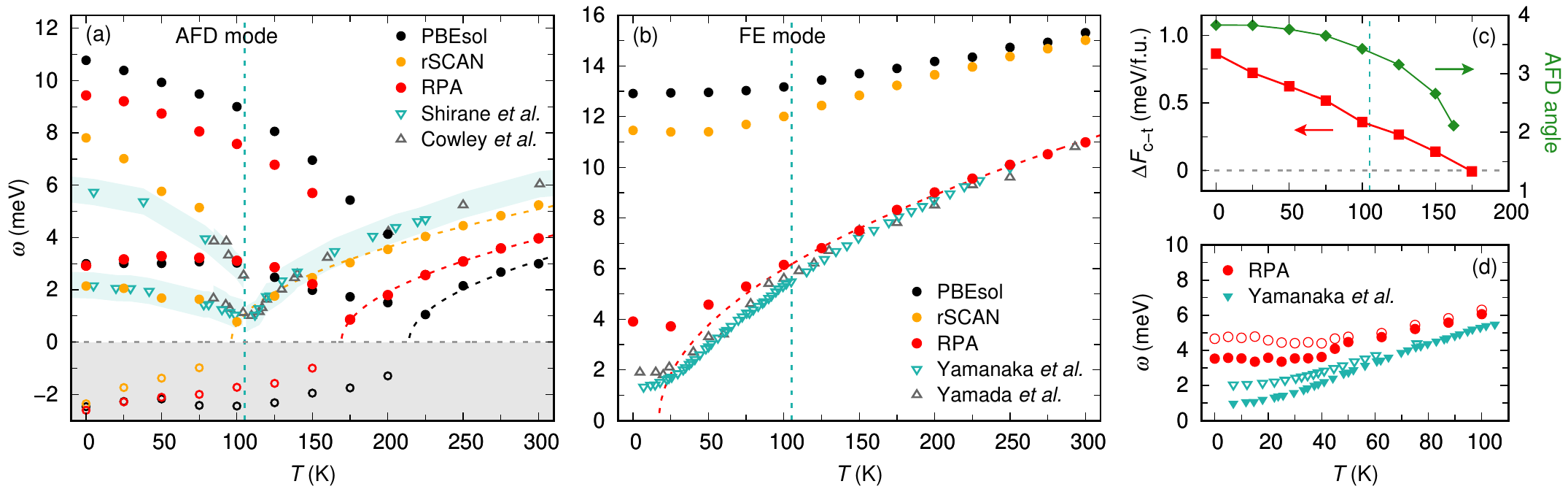}
 \caption{ \label{fig3}
 AFD phase transition and quantum paraelectricity in \STO. 
 (a) Temperature dependence of the AFD mode 
 calculated using MLFFs trained on PBEsol (black), rSCAN (orange) and 
 the RPA (red). The empty symbols in the negative frequency region 
 represent imaginary phonons obtained when the symmetry of the 
 structure is kept cubic at each temperature, and the shaded regions are 
 a guide to the eye. 
 (b) Temperature dependence of the FE mode calculated using the same 
 MLFFs. When the structure is tetragonal, averages over the $A_{2u}$ 
 and $E_u$ modes are shown. 
 In (a) and (b), experimental measurements from Refs.~\cite{Shirane1969,
 Cowley1969,Yamanaka2000,Yamada1969} are also reported. The dashed 
 vertical lines indicate the experimental AFD transition temperature, 
 and the dashed curves are Curie-Weiss fits.
 (c) Free-energy difference between the cubic and tetragonal structures 
 and AFD rotation angle in the tetragonal phase as a function of 
 temperature obtained from the RPA. Above 175~K, the most stable 
 structure is the cubic one. 
 (d) Close-up of the FE soft mode in the low-temperature region from 
 the RPA and hyper-Raman spectroscopy~\cite{Yamanaka2000}, showing 
 the splitting into the $A_{2u}$ (empty symbols) and $E_u$ (filled 
 symbols) branches. }
\end{center} \end{figure*}

We now proceed to include anharmonicity by combining SSCHA and 
MLFFs. 
Figure~\ref{fig2} displays the anharmonic phonon dispersions of 
tetragonal and cubic \STO\ at different temperatures, computed at 
the RPA level from the free-energy Hessian~\cite{Monacelli2021,Note1}. 
Here, for each temperature we adopt the experimental lattice 
volume, and we allow the $c/a$ ratio to relax so as to minimize the 
temperature-dependent anharmonic free energy. The corresponding 
harmonic dispersions are also shown. 
As expected, in the harmonic approximation the cubic structure exhibits 
unstable phonon modes with imaginary frequencies at the $\Gamma$ 
and R points. 
The latter is responsible for the low-temperature AFD transformation 
into the tetragonal structure, and indeed it remains unstable below 
the transition temperature even when anharmonic effects are included 
[see the arrow in Fig.~\ref{fig2}(a)]. In the tetragonal structure, this 
triply degenerate mode becomes a zone-center phonon and splits into 
an $A_{1g}$ mode, corresponding to oxygen rotations around the $c$ 
axis, and a doubly degenerate $E_g$ mode, both found to be stable. 
On the contrary, in both structures the FE modes at $\Gamma$ are 
stable at every temperature when quantum anharmonic fluctuations 
are accounted for. This confirms the quantum paraelectric ground state 
of \STO. Similar conclusions apply for PBEsol and rSCAN, as reported 
in SM Fig.~S9~\cite{Note1}. 

Both the AFD and FE modes show a marked temperature dependence. 
As can be seen in Fig.~\ref{fig3}(a), the anharmonic phonon 
frequency of the AFD mode in the high-temperature cubic phase 
shows the softening characteristic of 
displacive phase transitions. The large splitting of the $A_{1g}$ 
and $E_g$ modes below the transition temperature $T_c$ is in line with 
the measured data~\cite{Shirane1969,Cowley1969}, with the $A_{1g}$ 
mode higher in energy than the $E_g$ one. A square-root Curie-Weiss 
fit yields values of $T_c$ of 214~K for PBEsol calculations, 96~K 
for rSCAN, and 172~K for the RPA. In Fig.~\ref{fig3}(c) the RPA 
free energy difference between the cubic and tetragonal phase is 
shown, as well as the AFD rotation angle, in line with the predicted 
$T_c$ from the phonon collapse. While none of the methods accurately 
reproduces the experimental $T_c$ of 105~K, rSCAN is in fairly good 
agreement. We remark that no other study has reported a more accurate 
value of $T_c$ without adjusting for some finite pressure. We find 
that this result is linked to the superior performance of rSCAN in 
describing the equilibrium tetragonal structure, in particular 
the AFD rotation angle and $c/a$ ratio (see SM Table~SI~\cite{Note1}). 
The RPA slightly overestimates these parameters as well as the 
equilibrium volume. Correspondingly, the energy gain following the 
AFD rotations and lattice elongation is slightly too large, resulting in 
a higher $T_c$. This is related to the reduced accuracy of the RPA 
in the description of short-range interactions involved in covalent 
bonds~\cite{Scheffler2012,Ruzsinszky2021}, such as the Sr-O bond that 
is mainly responsible for the AFD instability~\cite{Spaldin2014}. 
Note that for all levels of theory, 
the effect of anharmonicity is to \textit{decrease} the equilibrium 
value of $c/a$ and the AFD rotation in the tetragonal phase (see SM 
Table~SI~\cite{Note1}). Thus, anharmonicity partly cures 
the so-called super-tetragonality problem in describing the structural 
properties. 

Moving to the FE instability, we have already shown that anharmonic 
quantum fluctuations suppress it down to 0~K. We remark that this 
result does not adopt any model assumptions nor does it suffer from 
the limitations of path-integral based methods at very low temperatures, 
where these methods become almost inapplicable~\cite{Vanderbilt1996}. 
This is precisely the temperature range where the onset 
of quantum critical effects is observed. 
However, the question remains of how well the FE soft mode 
is described by first-principles calculations as compared to the 
experimental measurements. From Fig.~\ref{fig3}(b) we see that both 
PBEsol and, surprisingly, rSCAN fail to reproduce the very soft 
experimental frequencies and their temperature dependence. The RPA 
comes very close to experiment, reproducing the plateau below around 
25~K as well as the subsequent increase with temperature.  
Correspondingly, in SM Fig.~S11~\cite{Note1} we also show that the 
calculations display the quantum critical scaling of the dielectric 
function observed experimentally. The energy 
splitting of the $E_u$ and $A_{2u}$ FE modes is also correctly captured, 
as shown in Fig.~\ref{fig3}(d). The sign and value of the splitting 
are dictated by the anharmonic relaxation of the $c/a$ ratio (see 
SM Fig.~S10(d) and the related discussion~\cite{Note1}). This effect is 
not generally taken into account, however it is tightly linked to the 
strain-induced ferroelectricity and the onset of quantum criticality.

Since the FE instability is highly sensitive to the lattice volume 
and shape, as seen in Figs.~\ref{fig1}(c) and \ref{fig1}(f), we 
consider the effect of anharmonic lattice expansion using rSCAN. 
After including this effect, the FE mode frequencies decrease only by about 
2~meV as shown in SM Fig.~S10(b)~\cite{Note1}, hence the experimental values 
are still largely overestimated. Additional calculations using 
the hybrid functional HSE06 also yield too hard FE anharmonic peaks 
(8~meV at 0~K). This confirms that the FE instability and its 
associated well depth, shown in Figs.~\ref{fig1}(a) and \ref{fig1}(b), 
can only be described accurately using the RPA, which gives a 
better description of the Ti-O interactions mainly 
responsible for the FE instability~\cite{Kresse2008}.
We argue that this conclusion is rather general and applies to other 
prototypical ABO$_3$ perovskites such as KTaO$_3$ and BaTiO$_3$. 
In the case of KTaO$_3$, anharmonic calculations using SCAN 
qualitatively reproduced the quantum paraelectric behavior, 
but also yielded too large FE frequencies~\cite{Ranalli2022}. Based on our 
results for SrTiO$_3$ and on the calculated FE well depths shown in 
Fig.~\ref{fig1}(g) for KTaO$_3$, we expect that the RPA 
would again produce softer frequencies in line with experiments. In 
contrast, HSE06 would predict even larger 
FE frequencies.  
BaTiO$_3$ undergoes a transition from a paraelectric cubic 
phase to a ferroelectric tetragonal one near 400~K. 
The transition temperature predicted from DFT 
is too low~\cite{Vanderbilt2002,Gigli2022}, implying 
that the energy barrier for the transition is underestimated. Our 
calculations for BaTiO$_3$ indicate that the increased well depths 
in the RPA should restore agreement with experiment 
(see Fig.~\ref{fig1}(h) and SM Fig.~S12~\cite{Note1}).

In conclusion, we use machine-learned force fields to perform 
non-perturbative anharmonic calculations to investigate the 
quantum paraelectric state in \STO\ as well as its AFD transformation 
between a cubic and tetragonal structure. Our calculations are free 
from commonly used approximations and are able to accurately 
characterize the temperature-dependent soft phonon modes. 
By leveraging $\Delta$-machine learning, we perform RPA-level calculations 
and show that the RPA predicts soft FE modes in good agreement with 
the experiment, whereas the most accurate semilocal functionals 
available do not. We also find that the RPA overestimates the 
AFD transition temperature, indicating that even beyond-RPA 
schemes are needed to cure the tendency of the RPA to 
underbind and overestimate bond lengths and volumes~\cite{Scheffler2012,Ruzsinszky2021}.
More generally, our work challenges the ability of DFT-based methods 
to describe key phenomena in quantum paraelectrics and 
conventional ferroelectrics, namely, dilute superconductivity and 
quantum criticality~\cite{Balatsky2015,Coleman2022}, non-linear 
phononics processes~\cite{Nelson2019,Hoffmann2019}, 
and electron-phonon coupling effects~\cite{Bernardi2018,Zacharias2020}. 
The approach we demonstrated is general and opens up possibilities 
to investigate materials exhibiting strong anharmonicity in combination 
with other unconventional quantum mechanical properties, such as 
many perovskite structures, lead and tin chalcogenides, and half-Heusler 
compounds. 

\begin{acknowledgments}
 C.V. and G.K. gratefully acknowledge many discussions with R. Jinnouchi, 
 F. Karsai and P. Liu. L.R. and C.F. thank L. Monacelli and M. Calandra 
 for helpful suggestions. The computational results have been mainly 
 achieved using the Vienna Scientific Cluster (VSC). This work was 
 supported by the Austrian Science Fund (FWF) within project SFB TACO 
 (Project No. F 81-N). 
\end{acknowledgments}

%

\end{document}